\documentclass[journal]{IEEEtran}

\usepackage{graphicx}
\usepackage{dcolumn}
\usepackage{bm}
\usepackage{float}
\usepackage{amsmath}
\usepackage{cite}

\ifCLASSINFOpdf
\else
\fi
\begin{document}
%
\title{A New Strategy for Designing Dual-band Antennas Based on Double-layer Metasurfaces}
%
%
%

\author{Kristy Hecht, Nacer Chahat, ~\IEEEmembership{Fellow,~IEEE}, Goutam  Chattopadhyay,~\IEEEmembership{Fellow,~IEEE}, Enrica Martini,~\IEEEmembership{Senior Member,~IEEE}, and Mario Junior Mencagli,~\IEEEmembership{Senior Member,~IEEE}
\thanks{The research was carried out at UNC Charlotte in collaboration with the Jet Propulsion Laboratory, California Institute of Technology, under a contract with the National Aeronautics and Space Administration.}
\thanks{Kristy Hecht and Mario Junior Mencagli are with the Department of Electrical and Computer Engineering, University of North Carolina at Charlotte, Charlotte, North Carolina 28223, USA (e-mail: 
\{khecht2, mmencagl\}@uncc.edu).}
\thanks{Nacer Chahat and Goutam Chattopadhyay are with Jet Propulsion Laboratory (JPL), California Institute of Technology, Pasadena, CA 91109, USA (e-mail: 
\{nacer.e.chahat,goutam.chattopadhyay\}@jpl.nasa.gov).}
\thanks{Enrica Martini is with the University of Siena, Department of Information Engineering and Mathematics, 53100 Siena, Italy (e-mail: 
martini@dii.unisi.it).}}

%
%

\markboth{Submitted to IEEE TRANSACTIONS ON ANTENNAS AND PROPAGATION on July 27, 2023}%
{Journal of \LaTeX\ Class Files,~Vol.~14, No.~8, August~2015}
%



\maketitle

\begin{abstract}
We present a new strategy for the design of dual-band planar antennas based on metasurfaces (MTSs) in the microwave and millimeter-wave regimes. It is based on a double-layer structure obtained by cascading two subwavelength patterned metallic claddings supported by a grounded dielectric slab. Each metallic layer is responsible for controlling radiation at one frequency, and it is engineered to be transparent at the other frequency. 
Hence, the two metallic layers can be designed independently using the well-established techniques for the design of single-layered MTS antennas. Layers decoupling is achieved by suitably exploiting Foster's reactance theorem, which regulates the frequency response of the impedance sheet modeling the metallic layer. By using the proposed strategy, a dual-band double-layered MTS antenna radiating a circularly polarized broadside beam in two popular frequency bands for climate science applications is designed and numerically verified.
\end{abstract}


%
\IEEEpeerreviewmaketitle

\section{Introduction} \label{sec1}
%
%
%
%
\IEEEPARstart{T}{he} demand for multifrequency medium-to-high gain antennas is rapidly increasing in many application areas, including Earth and climate science, remote sensing, satellite communications and 5G. These antennas, by exploiting the same radiating aperture at different frequency bands, offer the possibility to increase the channel capacity, improve the isolation between the transmitted and received signals, and provide multiple functionalities.\\
Reflectarrays \cite{chaharmir2010,guo2016,malfajani2014,han2005,hamzavi-zarghani2015,qu2014} and transmitarrays \cite{wu2017,matos2017,pham2017,pham2021,aziz2019,cai2020} have been largely explored as radiators operating in multiple frequency bands. Among other advantages, these antenna solutions are characterized by low cost, low weight, and ease of fabrication. However, they usually require an external feed, which makes them less appealing for applications with severe space constraints. Thus, it is of significant interest to propose multifrequency antenna solutions with an integrated feed.

Metasurface (MTS) antennas have raised a great deal of interest in recent years \cite{fong2010,patel2011,minatti2014,minatti2016,minatti2016_2,gonzalez2017,gonzalez2018_2, Casaletti}. Key features they offer include lightweight, ultra-thin form factor, polarization control, and a simple feeding mechanism embedded in the radiating aperture. At microwave and millimeter-waves, MTS antennas are typically realized with a layer of metallic patches arranged on top of a grounded dielectric slab. In the simplest cases, the feed structure is a vertical monopole placed inside the slab, at the center of the antenna and coaxially powered from the ground plane. This monopole excites a cylindrical surface wave that is gradually converted into a radiative wave, known as a leaky wave, through the interaction with the metallic texture, which features a periodic modulation. The metallic texture can be modeled by a spatially modulated surface impedance \cite{martini2020} which defines the boundary conditions between the tangential components of the electric and magnetic fields. Under the hypothesis of negligible losses (which is reasonable in the microwave and millimeter-wave regimes) this surface impedance can be assumed to be purely imaginary (reactive). The corresponding equivalent reactance must be a monotonically increasing function of the frequency alternating poles and zeros, as stated by the Foster’s reactance theorem \cite{Foster, maci2015}. 

In the present work, the behavior of the MTS surface reactance versus frequency is exploited to design dual-frequency MTS antennas. Prior works on dual-band MTS antennas are typically based on two different strategies. The first one consists of superimposing two different surface impedance modulations, each one responsible for controlling radiation at one frequency \cite{li2018,bodehou2020,faenzi2021}. The metasurface implementation, however, results quite challenging for radiative systems operating at two largely separated frequency bands, due to the conflicting requirements on the unit cell size at the two frequencies.
The second strategy is based on the external illumination of two cascaded surface impedances, whose profiles are obtained through an optimization \cite{budhu2022}. While this approach potentially has no restriction on the separation between the two operating frequencies, the presence of an external excitation results in a poor device form factor. Our proposal allows overcoming such issues by combining the benefits of the two strategies. It consists of a dual-layered MTS antenna capable of performing at two broadly different frequencies with a coplanar excitation. The key point in our proposal is to make the layer working at the frequency $f_1$ transparent at the frequency $f_2$, and vice-versa. As mentioned above, this behavior can be achieved by exploiting the Foster’s reactance theorem. Namely, the equivalent impedance of the metallic layer operating at $f_1$ ($f_2$) has to be close to a pole at $f_2$ ($f_1$). 

After briefly reviewing the design principle of single-layered MTSs operating in one frequency band, Sec.~\ref{sec2} discusses how to build on the Foster's reactance theorem to design double-layered MTS antennas capable of performing in two different frequency bands. Sec.~\ref{sec3} presents a realistic implementation of the proposed approach by designing a double-layered MTS antenna radiating circularly polarized broadside beams at two common frequencies in cloud and precipitation radar \cite{nagaraja2021}, such as $35.75$GHz and $94.05$GHz. The performance of the antenna is verified by full-wave simulation. Finally, we draw our conclusions in Sec.~\ref{sec4}.

\section{Conceptual Model} \label{sec2}
\subsection{Single-layered MTS antennas} \label{subsec2A}
To begin with, we briefly review the basic principle behind the design of single-layered MTS antennas. Then, building upon this principle, we will discuss our approach to designing double-layered MTS antennas for dual-frequency operation. An $e^{j\omega t}$ time convention is used and suppressed throughout the paper.
\begin{figure}[ht]
\centering
\includegraphics[width=0.98\columnwidth]{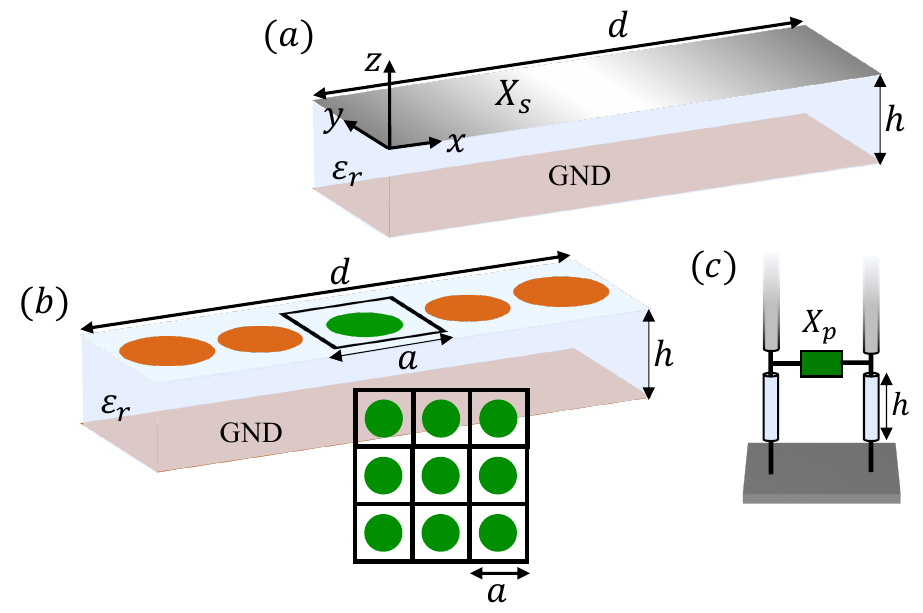}
\caption{(a) A 2-D canonical problem consisting of a penetrable sinusoidally modulated surface reactance ($X_s$) on top of a grounded dielectric slab. The geometry is assumed to be invariant along the $y$-direction and periodically modulated along the $x$-direction. (b) $X_s$ is implemented through gradually varying subwavelength patches. The inset shows the locally periodic problem utilized to map the patches into an equivalent reactnce ($X_p$) through (c) the local transmission line model.} 
\label{two}
\end{figure}

The design of single-layered MTS antennas is based on the two-dimensional (2-D) canonical problem shown in Fig.~\ref{two}(a). It consists of a periodically modulated sheet reactance lying on a dielectric substrate backed by a metallic ground plane. The structure is assumed to be invariant along the $y$ direction and we consider propagation along $x$. $\varepsilon_r$ and $h$ denote the relative permittivity and the thickness, respectively, of the dielectric slab. The sheet reactance imposes the following boundary condition 
\begin{equation}\label{IBC}
\boldsymbol{E}_t\left( x,z=0 \right)=jX_s\left( x \right)\left [ \boldsymbol{H}_t\left( x,z=0^{+} \right) - \boldsymbol{H}_t\left( x,z=0^{-} \right) \right]
\end{equation}
and its modulation profile is usually described by a sinusoidal function
\begin{equation}\label{eq1}
Z_s\left( x \right) = jX_s\left( x \right) = jX_0 \left [ 1+M \text{cos}\left(\frac{2\pi x}{d}\right) \right]
\end{equation}
where $\boldsymbol{E}_t$ and $\boldsymbol{H}_t$ are the tangential components of the electric and magnetic fields, respectively, at the sheet level, $X_0<0$ is the average reactance, $M$ is the modulation index, and $d$ is the modulation period, . This structure supports a TM (transverse magnetic) mode without cut-off. Due to the periodic nature of $X_s$, such a mode can be represented through an infinite set of Floquet waves (FWs) \cite{oliner1959,martini2020}. The longitudinal (along $x$) wavenumber of the $n$-indexed FW is
\begin{equation}\label{eq2}
k_{x}^{\left(n\right)} = k_{x}^{\left(0\right)}+\frac{2\pi n}{d}
\end{equation}
where $k_{x}^{\left(0\right)}$ is the longitudinal wavenumber of the $0$-indexed  FM, which is solution of the dispersion equation. For $d$ larger than a critical value \cite{oliner1959}, at least one of the higher-order FMs ends up inside the light cone, and leaky waves occur. In this case, the solution of the dispersion equation takes a complex value and can be represented as 
\begin{equation}\label{eq3}
k_{x}^{\left(0\right)} = \beta_{sw}+\Delta \beta - j\alpha
\end{equation}
where $\beta_{sw}$ is the longitudinal wavenumber of the surface wave propagating on the unmodulated surface impedance ($M=0$), and the positive real numbers $\Delta \beta$ and $\alpha$ account for the perturbation in the phase and amplitude, respectively, induced by the modulation, with $\Delta \beta \ll \beta_{sw}$. In MTS antennas, the modulation period ($d$) is usually chosen such that only the ($-1$)-indexed FM is inside the light cone. As a result, the pointing angle ($\theta$), which
is defined with respect to the z axis, and $d$ are related as
follows
\begin{equation}\label{eq4}
\beta_{sw}+\Delta \beta - \frac{2\pi}{d}= k \text{sin}\theta
\end{equation}
with $k$ being the free-space wavenumber.

As mentioned in Sec.~\ref{sec1}, the desired continuous impedance profile can be accurately synthesized by subwavelength metallic patches smoothly varying along the surface \cite{martini2020} (see Fig.~\ref{two}(b)). The gradual variation of the patches allows for the use of the local periodicity approximation. Namely, the equivalent impedance of each patch can be extracted as if it was embedded in a periodic environment [inset of Fig.~\ref{two}(b)]. A periodic pattern modeled with the equivalent transmission line model shown in Fig.~\ref{two}(c) is studied to build databases linking the equivalent impedance ($Z_p = jX_p$) with the geometrical parameters of the unit cell. Different techniques for an efficient and accurate impedance extraction of periodic metallic claddings printed on a grounded slab are available in the published literature (see, e.g.,\cite{luuk2008,mencagli2015,mencagli2016}).

\subsection{Dual-band Double-layered MTS antenna} \label{subsec2B}
Based on the procedure summarized in the previous section, our goal now is to develop a new methodology that allows performing two independent MTS antenna designs working at two different frequencies, that can be merged together, resulting in a single flat dual-band radiator. We assume the two frequencies of interest are $f_1$ and $f_2$, with $f_1<<f_2$. Henceforth, all the frequency-dependent physical quantities introduced in the previous section are assigned with subscripts $1$ and $2$, corresponding to $f_1$ and $f_2$, respectively. 

\begin{figure}[ht]
\centering
\includegraphics[width=0.98\columnwidth]{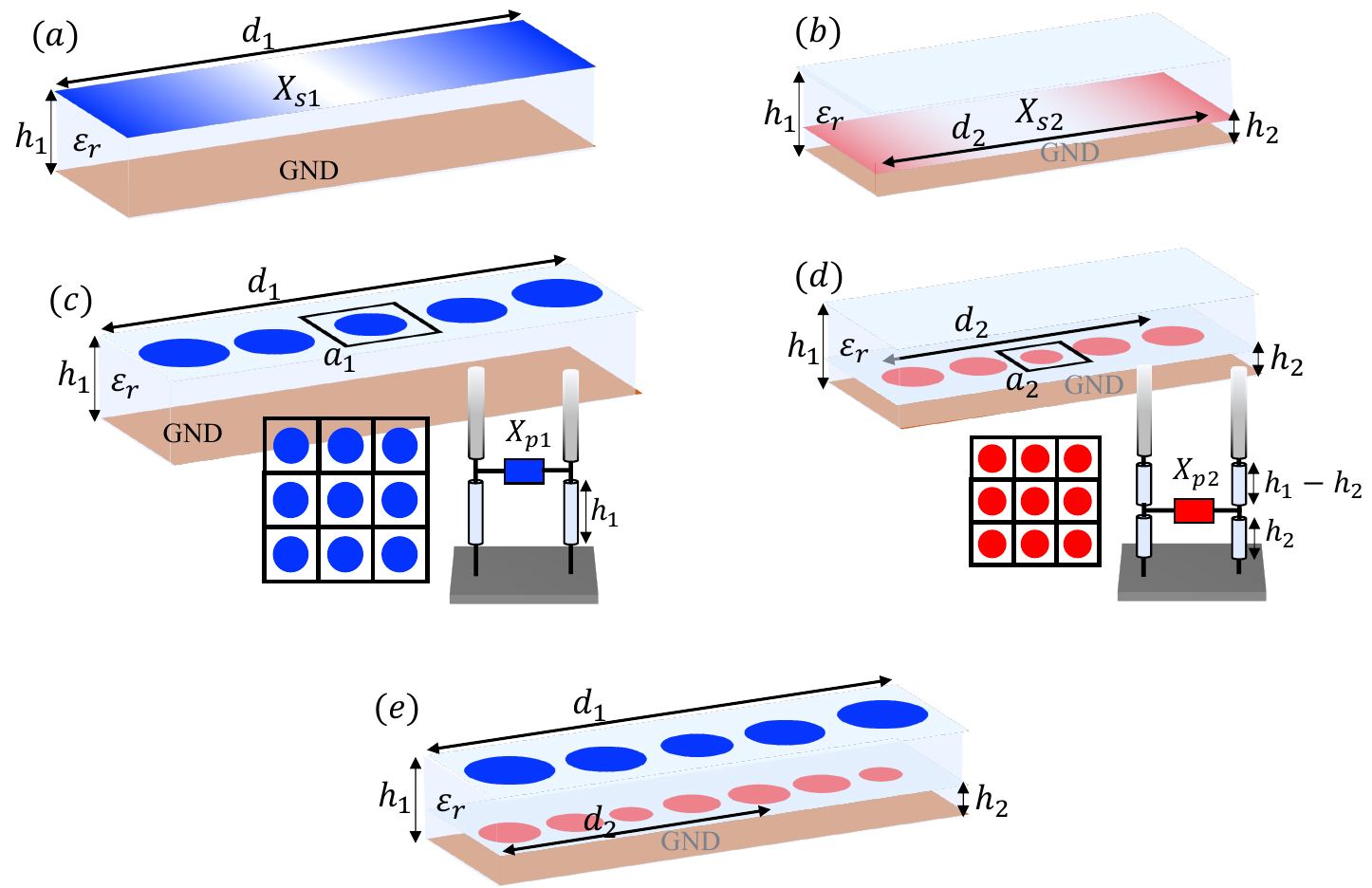}
\caption{(a) and (b) 2-D canonical problems operating at $f_1$ and $f_2$, respectively.  (c) and (d) implementation with metallic patches. The insets in (c) and (d) show the local periodic problem and transmission line model. (e) Combination of the structures in (c) and (d) performing at $f_1$ and $f_2$.} 
\label{three}
\end{figure}

Let us assume we have the two canonical problems shown in Figs.~\ref{three}(a) and (b). The first one [Fig.~\ref{three}(a)], which is identical to that shown in Fig.~\ref{two}(a) with a surface reactance sheet $X_{s1}$ and slab thickness $h_1$, is set up to perform at $f_1$ with a pointing angle $\theta_1$. The second one [Fig.~\ref{three}(b)] is slightly different than the previous one. That is, the sinusoidally modulated reactance sheet ($X_{s2}$) is embedded in the dielectric slab and located at a distance $h_2$ from the ground plane, with $h_2<h_1$. This difference in the stack-up is rigorously accounted for by properly defining the problem's Green's function. Hence, the antenna design principle remains unchanged and the modulation is set up to operate at $f_2$ with a pointing angle $\theta_2$. As discussed in Sec.~\ref{subsec2A}, the two reactance sheets can be implemented by gradually changing the size and geometry of subwavelength patches [see Figs.~\ref{three}(c) and (d)]. Each patch is linked to an equivalent impedance extracted from an equivalent transmission line model based on the local periodicity approximation. This model can be defined in the two cases as shown in the insets of Figs.~\ref{three}(c) and (d), respectively, with $X_{p1}$ and $X_{p2}$ representing the equivalent reactance of the patches in a locally periodic environment. 

Now, given the two MTSs of Figs.~\ref{three}(c) and (d) operating at $f_1$ and $f_2$, respectively, we wish that their combination [Fig.~\ref{three}(e)] is capable of performing at both frequencies of interest. At first sight, the obtained double-layered MTS cannot perform equally well as the two single-layered MTSs. In fact, when the two metallic patterns are placed closely one to another, their coupling through near-field interactions is likely to imply a change in the equivalent impedance of each pattern. Since the patterns have been designed independently, the radiation performances can be noticeably affected by this change. In principle, the coupling between the two metallic layers can be rigorously accounted for in the design process by extracting the equivalent reactance of the patches in the two-layer configuration. However, this would results in a procedure significantly more complicated than the well-established techniques available for the design of MTSs \cite{luuk2008,mencagli2015,mencagli2016}. Also, in order this approach to be applicable, it is necessary to impose a restriction on the size of the unit cells for the two lattices: $a_1$ and $a_2$ must be commensurable, so that a common period could be identified. 

An alternative approach that does not require the extraction of the equivalent impedance in double-layered periodic structures and does not pose any restriction on the unit cell sizes would be preferable. The proposed approach aims at satisfying these requirement by focusing on a suitable impedance synthesis of the two single-layered MTSs [Figs.~\ref{three}(c) and (d)] such that they perform equally well when they are combined together [Fig.~\ref{three}(e)].  
\begin{figure}[ht]
\centering
\includegraphics[width=0.98\columnwidth]{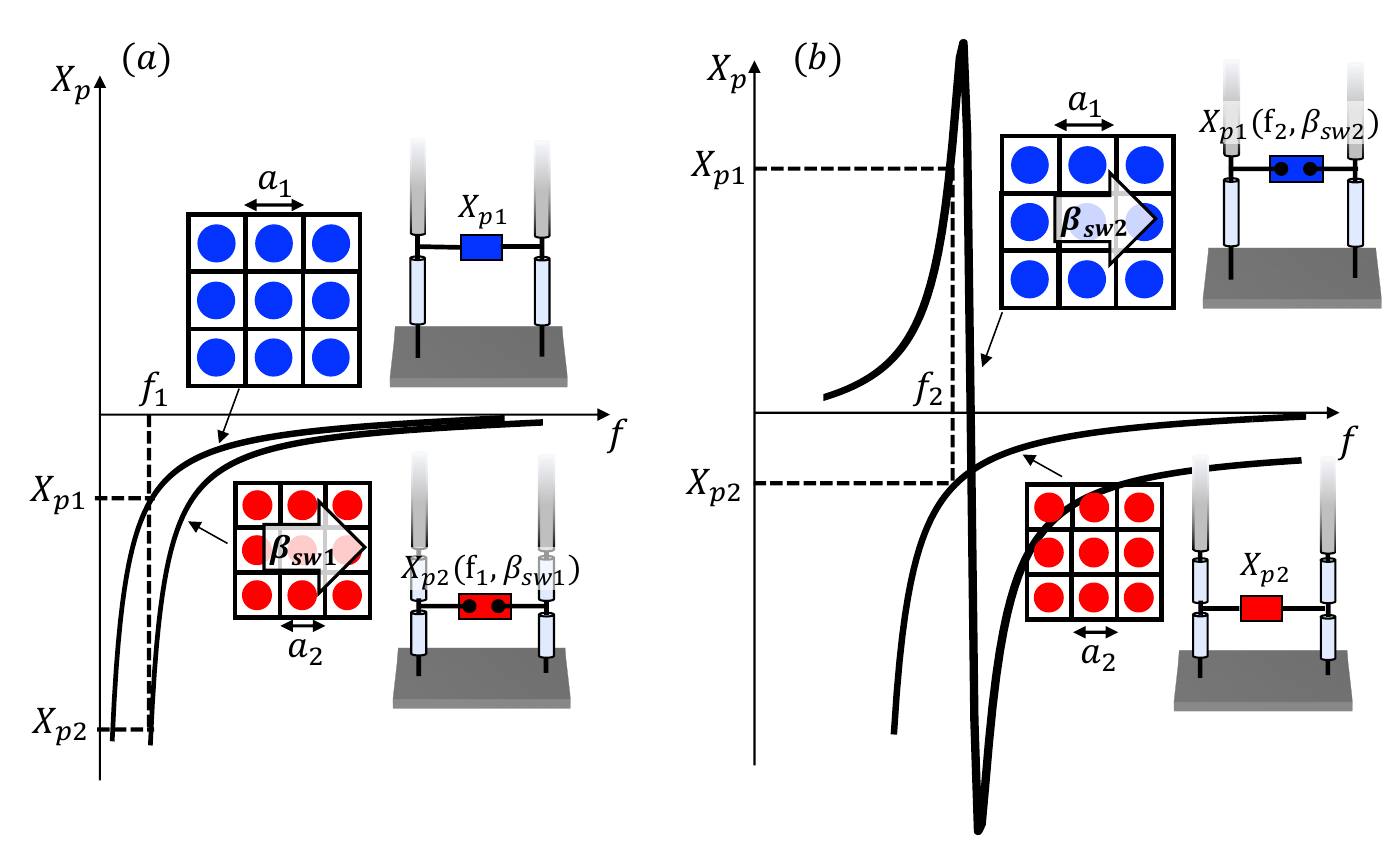}
\caption{(a) Frequency dependence of the equivalent reactance resulting from the local periodic problem (top inset) of the patches working at $f_1$ (left curve). The right curve shows the equivalent impedance of the patches designed to work at $f_2$ (bottom inset) seen from the wave supported by the locally periodic structure in the top inset. (b) The same as in (a), but around $f_2$. The resonant curve represents the equivalent reactance of the patches designed to work at $f_1$ (bottom inset) seen from the wave supported by the locally periodic structure in the bottom inset.} 
\label{four}
\end{figure}

As is well known, the equivalent reactance provided by a periodic array of metallic patches is purely imaginary and exhibits a capacitive behavior in the low frequency regime. As a result, according to the Foster's reactance theorem, its frequency response presents a pole at zero frequency and then it increases monotonically alternating zeros and poles for increasing frequencies. The unit cells of single-frequency MTS antennas are normally designed to operate between the first zero and the first pole. Hence, we assume $X_{p1}$ and $X_{p2}$ also reside in that region at their operative frequency
(see $X_{p1}$ and $X_{p2}$ in Figs.~\ref{four}(a) and (b), respectively). Now, the question is: what is the equivalent reactance of the patches designed to operate at one frequency as seen from the wave supported by the single-layered MTS operating at the other frequency? To address this question, let's start from the patches designed to operate at $f_2$ (red patches in Fig.~\ref{three}(e)). Since our objective is not to perturb the behaviour of the impedance sheet $X_{p1}$ at $f_1$, we assume to have at that frequency the same wave that would be supported by the single-layered MTS [Fig.~\ref{three}(c)], corresponding to having an infinite reactance (open circuit) in the bottom layer. Hence, to verify the consistency of the assumption, we need to evaluate the equivalent impedance of the patches implementing the profile $X_{p2}$ by simultaneously setting the frequency equal to $f_2$ and the longitudinal wavenumber equal to $\beta_{sw1}$ (see the bottom inset of Fig.~\ref{four}(a)). Notice that this combination does not satisfy the resonance equation, and therefore it does not correspond to a mode supported by the structure. 
Hence, this process implies extracting the reactance of the patches {\it outside the relevant dispersion curve} and can be carried out through, for example, the periodic Method of Moment (MoM) described in \cite{maci2006}.
In practice, since we assumed $f_1<<f_2$, $a_2$ will be much smaller than $a_1$ and so will be the relevant patches. Thus, we can assume that their equivalent reactance evaluated at $f_1$ and $\beta_{sw1}$ will be very close to the pole at zero frequency, corresponding to a quasi open-circuit [Fig.~\ref{four}(a)]. As a result, the MTS operating at $f_2$ [red patches in Fig.~\ref{three}(e)] are almost transparent at $f_1$, with no need for additional design strategies. 

The same procedure must be repeated to evaluate the equivalent reactance of the patches implementing $X_{p1}$ at $f_2$ and $\beta_{sw2}$. However, now $a_1$ and the relevant patches are electrically large at $f_2$. For this reason, the patches  working at $f_1$ [blue patches in Fig.~\ref{four}(b)] need to be properly designed such that their equivalent impedance presents a pole close to $f_2$, thus, behaving like a quasi-open circuit at that frequency [Fig.~\ref{four}(b)]. Note that this pole will be at higher frequency with respect to the first zero; depending on the separation between $f_1$ and $f_2$, it could be the second or the third pole. 

Following this strategy, the double-layered MTS of Fig.~\ref{three}(e) obtained from the combination of the two single-layered MTSs of Figs.~\ref{three}(c) and (d) operating at $f_1$ and $f_2$, respectively, is expected to perform well at both the frequencies of interest. 
\section{Antenna implementation} \label{sec3}
\begin{figure}[ht]
\centering
\includegraphics[width=0.98\columnwidth]{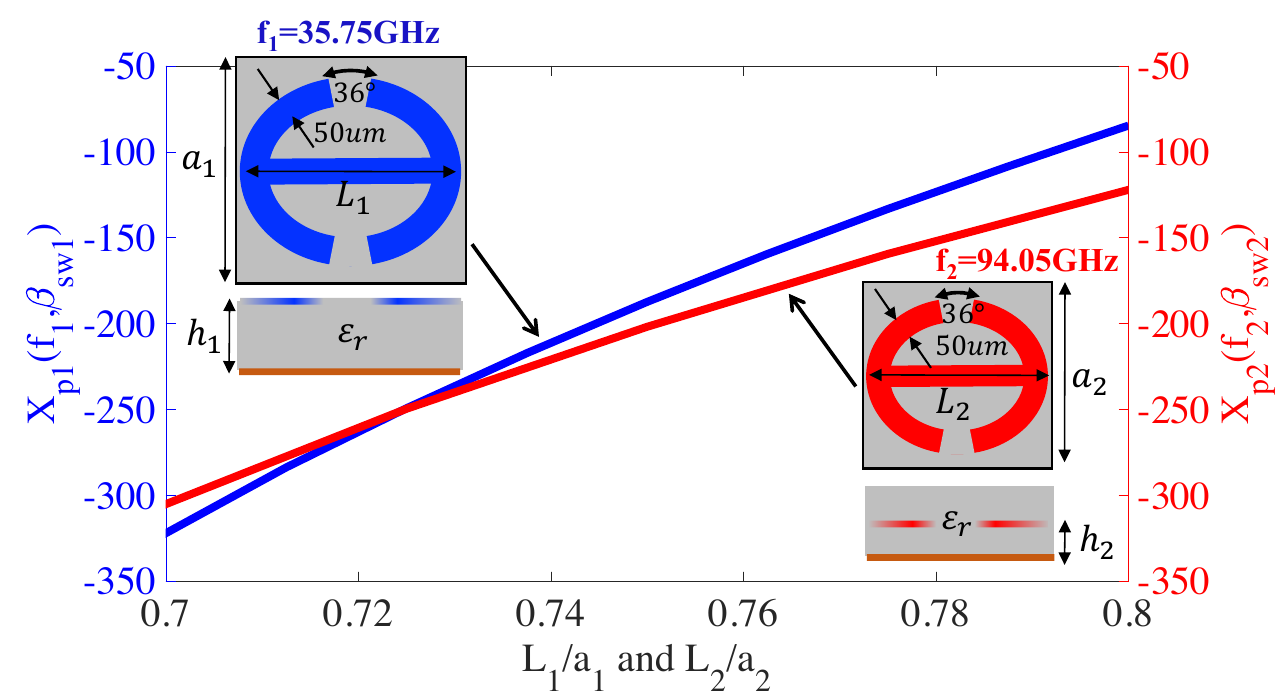}
\caption{Equivalent surface reactance versus the patch size for the unit cell operating at $35.75$GHz (top-left inset) and $94.05$GHz (bottom-right inset). $a_1$ and $a_2$ are $1.29$mm and $0.43$mm, respectively.} 
\label{five}
\end{figure}

\begin{figure}[ht!]
\centering
\includegraphics[width=0.98\columnwidth]{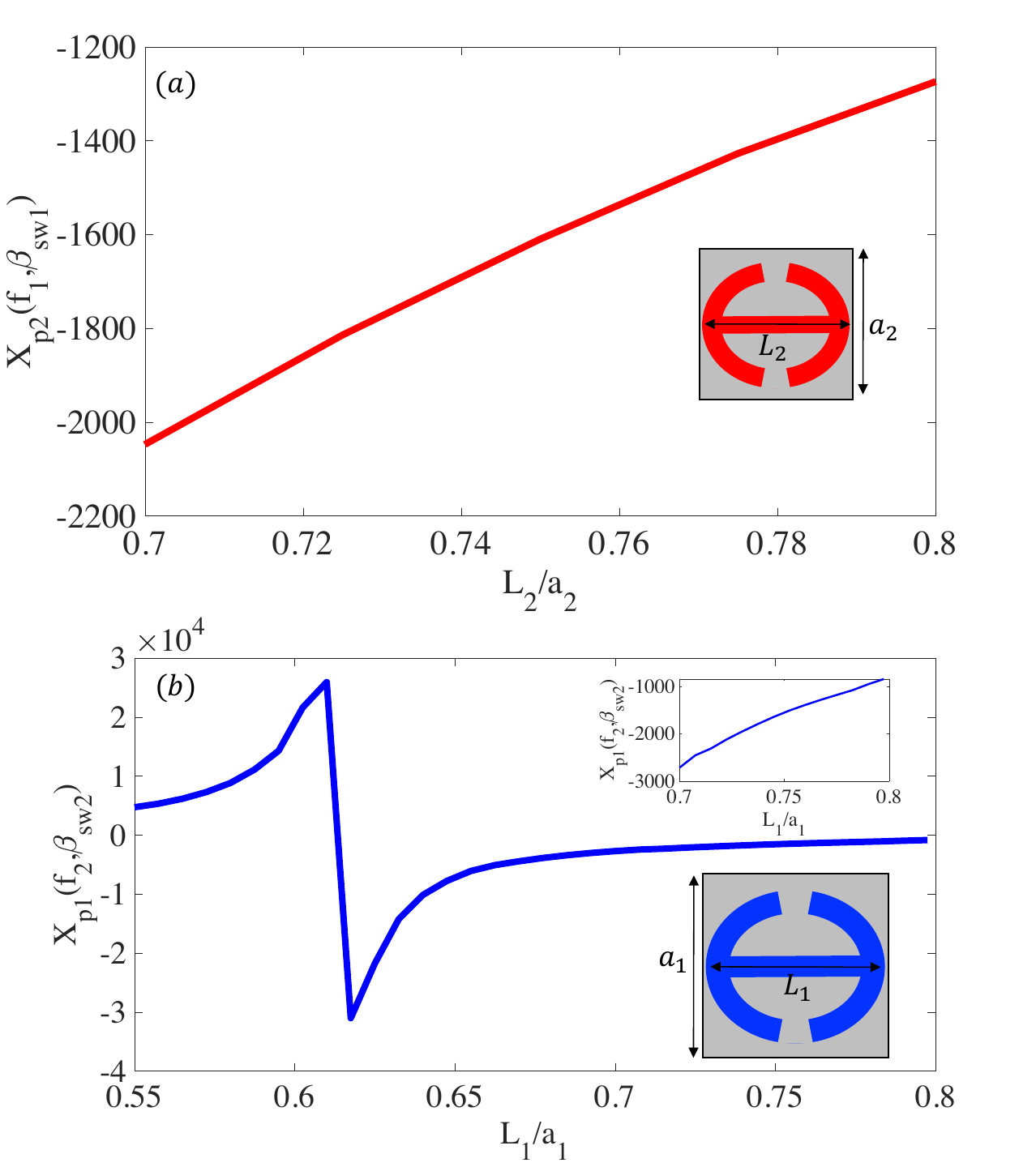}
\caption{Driven equivalent surface reactance versus the patch size. (a) The surface reactance is extracted by forcing a TM surface wave with frequency $35.75$GHz and wavenumber $\beta_{sw1}=777.8\,$rad/m in the local periodic problem with the unit cell operating at $94.05$GHz shown in the inset. (b) The same process as in (a) is repeated for the local periodic problem with the unit cell operating at $35.75$GHz shown in the bottom inset. The TM surface wave has the frequency $94.05$GHz and wavenumber $\beta_{sw2}=2981.27\,$rad/m. The top inset shows a zoom-in of the equivalent surface reactance with size of the patch ranging as in Fig.~\ref{five}.} 
\label{six}
\end{figure}

To demonstrate the feasibility of the proposed approach for the design of dual-band double-layered MTS antennas, we chose two largely separated popular frequencies in climate science radars such as $35.75$GHz ($f_1$) and $94.05$GHz ($f_2$) \cite{nagaraja2021}. The dielectric substrate is Roger $4003$C with $\varepsilon_r=3.55$ and thickness $h_1=406 \mu$m. The canonical problems at two frequencies [Figs.~\ref{three}(a) and (b)] were studied with the approach proposed in \cite{martini2020}. For the canonical problem at $f_2$ [Fig.~\ref{three}(b)], the Green's function in \cite{martini2020} was suitably changed to consider that the modulated reactance sheet sits inside the dielectric slab ($h_2=h_1/2$) instead of on the top. Defining $X_{s1}$ and $X_{s2}$ in accordance with the modulation profile in \ref{eq1} with $M_{1}=M_{2}=0.3$, $X_{01}=-150\Omega$, $X_{02}=-100\Omega$, $d_{1}=7.4$mm, and $d_{2}=2.15$mm, the canonical problems at $f_1$ and $f_2$ support a fundamental ($0$-indexed) TM FW with wavenumbers $k_{x1}^{\left(0\right)}=778.7-j0.08\,$rad/m and $k_{x2}^{\left(0\right)}=2985.5-j5.5\,$rad/m, respectively. With $M_{1}=M_{2}=0$ (unmodulated surface impedances), $k_{x1}^{\left(0\right)}$ and $k_{x2}^{\left(0\right)}$ reduce to $\beta_{sw1}=777.8\,$rad/m ($\Delta\beta_1=0.9$) and $\beta_{sw2}=2981.27\,$rad/m ($\Delta\beta_2=4.23$), respectively. With this setup, it is straightforward to verify through Eq.~(\ref{eq4}) that the ($-1$)-indexed FW at both frequencies resides inside the light cone generating a broadside beam ($\theta_1=\theta_2=0$). $X_{s1}$ and $X_{s2}$ are synthesized with the square unit cells shown in the top-left and bottom-right insets, respectively, of Fig.~\ref{five}. Both unit cells are based on the same metallic element topology, consisting of a double-anchor patch. The patch sits at the air-dielectric interface and within the dielectric slab ($h_2=h_1/2$) in the unit cell for $f_1$ and $f_2$, respectively (see the insets of Fig.~\ref{five}). The two unit cells with side $a_1=1.29$mm and $a_2=0.43$mm were studied with the MoM in \cite{maci2006} for different patch sizes forcing a longitudinal wavenumber $\beta_{sw1}$ at $f_1$ in the first case and a longitudinal wavenumber $\beta_{sw2}$ at $f_2$ in the second case. The reactances databases ($X_{p1}$ and $X_{p2}$) as a function of the patch size are shown in Fig.~\ref{five}. By using the same MoM, the equivalent TM reactance was extracted by forcing in the periodic problem with the unit cell designed for $f_2$ (bottom-right inset of Fig.~\ref{five}) a surface wave at frequency $f_1$ with longitudinal wavenumber $\beta_{sw1}$ [Fig.~\ref{six}(a)]. With this setup, one can observe that such a unit cell operates in extremely high impedance regime (quasi-open circuit) at $f_1$ for all the patch sizes of interest. The same process was repeated with the unit cell designed for $f_1$ (top-left inset of Fig.~\ref{five}) imposing a surface wave with frequency $f_2$ and longitudinal wavenumber $\beta_{sw2}$. The extracted reactance is shown in Fig.~\ref{six}(b). The range of $L_1$ was extended with respect to the one of Fig.~\ref{five} to show the presence of a pole. One can observe that the location of the pole is right before the range of operation assumed in the reactance database of Fig.~\ref{five}. Thus, the patches designed for $f_1$ operate in extremely high impedance regime (quasi-open circuit) at $f_2$ , as can be seen in the inset of Fig.~\ref{six}(b). It is worth emphasizing that, although in the design under consideration the relation between the sides of the two unit cells ended up being $a_1 = 3a_2$, our approach, as discussed in the previous section, does not prevent using unit cells whose sides are not commensurable. 
\begin{figure}[ht]
\centering
\includegraphics[width=0.98\columnwidth]{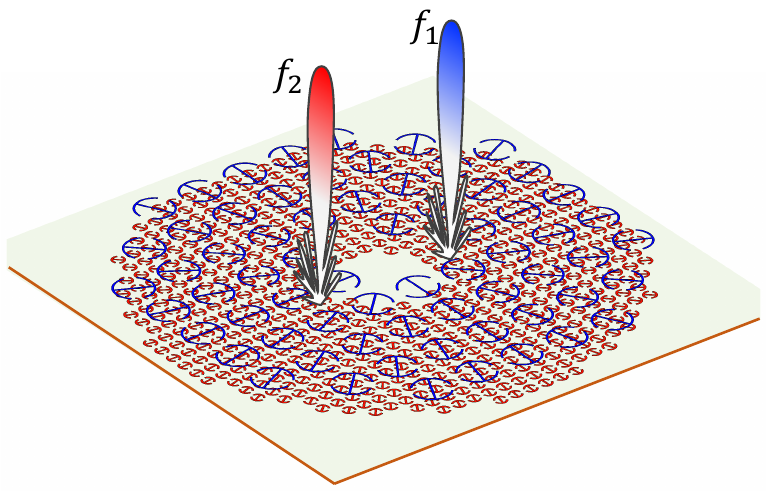}
\caption{Central portion of the layout of the designed double-layered metasurface antenna operating at $35.75$GHz and $94.05$GHz. The structure consists of two metallic claddings supported by a grounded dielectric slab. The top and bottom claddings (blue and red patches) control the radiation at $35.75$GHz and $94.05$GHz, respectively. The bluish and reddish beams conceptually sketch radiation patterns at $f_1$ and $f_2$, respectively.} 
\label{one}
\end{figure}
\begin{figure}[ht]
\centering
\includegraphics[width=0.98\columnwidth]{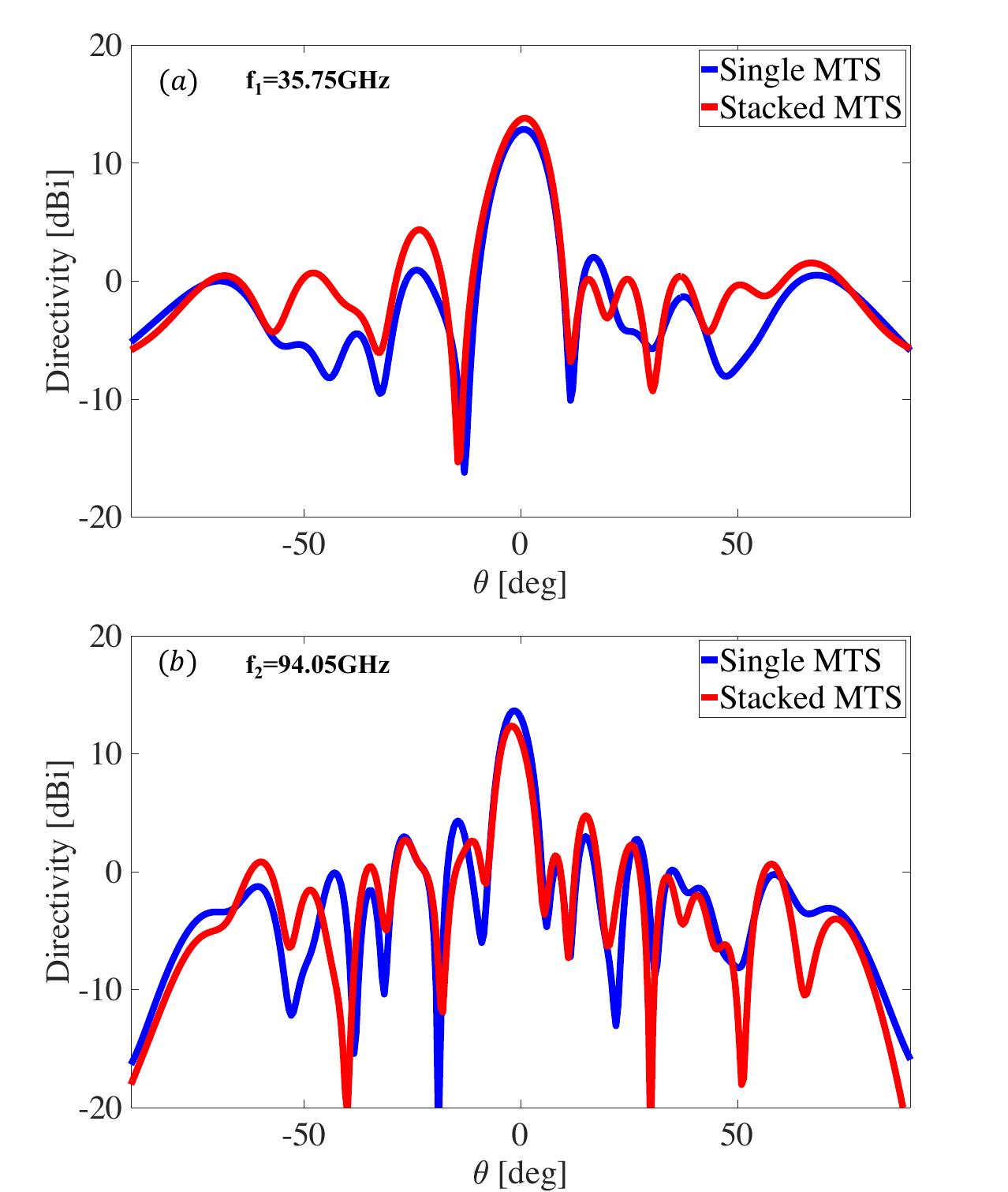}
\caption{Right-handed circularly polarized directivity patterns for the single- and double-layered MTS antenna of Fig.~\ref{one} at (a) $35.75$GHz and (b) $94.05$GHz.} 
\label{seven}
\end{figure}
By using the reactance databases of Fig.~\ref{five}, we designed a double-layered 3-D MTS antenna radiating a right-handed circular polarized (RHCP) wave at both frequencies. By 3-D, we mean a MTS where the metallic patches are modulated in two directions patterning the following modulated reactance sheets: $X_{s1,2}^{2D}\left( \rho,\phi \right) = X_{01,2} \left [1+M_{1,2} \text{cos}\left(\frac{2\pi }{d_{1,2}}\rho-\phi\right) \right]$ with $\rho$ and $\phi$ representing the position in polar coordinates. The $\phi$-dependence in $X_{s1,2}^{2D}$ ensures that any two sectors on the antennas separated by $90^{\circ}$ radiate an electric field with orthogonal and quadrature-phased components, resulting in a circularly polarized wave. Note that the design of 3-D MTS antennas is usually based on the 2-D canonical problem of Fig.~\ref{two}. Each sector can be seen as a 2-D canonical problem rotated by an angle $\phi$ around the $z$-axis. Hence, the approach proposed in this manuscript can be applied to designing both 2-D and 3-D dual-band double-layered MTS antennas.

The geometry of the designed antenna is illustrated in Fig.~\ref{one}; for ease of visualization, only the central portion of the antenna layout is displayed. The sizes of the two MTSs are equal, although the approach proposed in this manuscript does not prevent the use of MTSs with different sizes. The antenna radius is $3\lambda$ and $5\lambda$ at the operating frequency $f_1$ and $f_2$, respectively. We selected a relatively small radius at the lower frequency in order to be able to complete the full-wave simulation with reasonable accuracy with our computational resources at both frequencies. The antenna is fed from the center by an electrically small dipole immersed in the dielectric, which excites a cylindrical TM mode. The dipole is displaced by $h_2$ upward relative to the center of the ground place. The antenna was simulated with the commercial software Ansys HFSS. Figs.~\ref{seven}(a) and (b) show the obtained co-pol directivity patterns at $f_1$ and $f_2$, respectively, for the single- and double-layered MTSs. One can observe a satisfactory agreement between the patterns generated by the single- and double-layered MTSs at both frequencies.
\section{Conclusion} \label{sec4}
A new strategy for the design of dual-band double-layered MTS antennas is presented. The structure of the considered antenna solution consists of a cascade of two MTSs supported by a grounded dielectric slab. Each MTS consisting of a subwavelength metallic cladding controls the radiation at one frequency. The coupling between the two MTSs is drastically reduced by exploiting the Foster's reactance theorem that dictates the frequency behavior of the reactance sheets modeling the metallic layers. The equivalent reactance of the unit cell working at one frequency must be close to a pole at the other frequency and vice-versa. By doing so, the two MTSs are inactive at the other frequency and can be designed independently, avoiding complicated impedance extraction techniques and restrictions on the unit cell sides. A double-layered MTS radiating a circularly polarized broadside beam at $35.75$GHz and $94.05$GHz is designed and numerically tested. The directivity patterns of the single- and double-layered MTSs agree pretty well at both frequencies. Although the concept was presented and verified for a scalar impedance with a constant modulation index, it can be readily extended to the case of anisotropic impedances, and more sophisticated modulation profiles. The proposed dual-frequency antenna solution may find applications in Earth and climate science, remote sensing, and satellite communications.



%


\ifCLASSOPTIONcaptionsoff
  \newpage
\fi



%

\bibliographystyle{IEEEtran}
\bibliography{IEEEabrv,mybib}

%






\end{document}